\documentclass[reprint,superscriptaddress, amsmath, amssymb, aps, prx]{revtex4-2}

\usepackage{dcolumn}% Align table columns on decimal point
\usepackage{bm}% bold math
\usepackage{hyperref}% add hypertext capabilities
\usepackage{graphicx}
\usepackage[version=3]{mhchem} % Formula subscripts using \ce{}
\usepackage[separate-uncertainty=true, per-mode=symbol]{siunitx}
\usepackage{physics}

\begin{document}

\title{Terahertz chiral sub-wavelength cavities breaking time-reversal symmetry via ultra-strong light-matter interaction}
\author{Johan Andberger}
%\email{ajohan@phys.ethz.ch}
\affiliation{Institute of Quantum Electronics, ETH Z\"urich, Auguste-Piccard-Hof 1, 8093 Z\"urich, Switzerland}
\author{Lorenzo Graziotto}
%\email{lgraziotto@phys.ethz.ch}
\affiliation{Institute of Quantum Electronics, ETH Z\"urich, Auguste-Piccard-Hof 1, 8093 Z\"urich, Switzerland}
\author{Luca Sacchi}
%\email{lsacchi@student.ethz.ch}
\affiliation{Institute of Quantum Electronics, ETH Z\"urich, Auguste-Piccard-Hof 1, 8093 Z\"urich, Switzerland}
\author{Mattias Beck}
%\email{mattias.beck@phys.ethz.ch}
\affiliation{Institute of Quantum Electronics, ETH Z\"urich, Auguste-Piccard-Hof 1, 8093 Z\"urich, Switzerland}
\author{Giacomo Scalari}
%\email{scalari@phys.ethz.ch}
\affiliation{Institute of Quantum Electronics, ETH Z\"urich, Auguste-Piccard-Hof 1, 8093 Z\"urich, Switzerland}
\author{J\'er\^ome Faist}
%\email{jerome.faist@phys.ethz.ch}
\affiliation{Institute of Quantum Electronics, ETH Z\"urich, Auguste-Piccard-Hof 1, 8093 Z\"urich, Switzerland}

\begin{abstract}
We demonstrate terahertz chiral sub-wavelength cavities that break time-reversal symmetry by coupling the degenerate linearly polarized modes of two orthogonal sets of nano-antenna arrays using the inter-Landau level transition of a two-dimensional electron gas in a perpendicular magnetic field, realizing normalized light-matter coupling rates up to $\Omega_R/\omega_{\mathrm{cav}} = 0.78$ with a dispersion that is modified by the parasitic capacitive coupling between the orthogonal antennas. The deep sub-wavelength confinement of the nano-antennas means that the ultra-strong coupling regime can be reached even with a small number of carriers compared to Fabry-Perot cavities, making it viable to be used with a variety of 2D materials. The non-degenerate circularly polarized ground state was only obtained after carefully optimizing the optical design to minimize the parasitic coupling to linearly polarized light.
\end{abstract}

\maketitle

A new class of quantum materials possessing non-classical properties has emerged along with the growing understanding of how quantum effects control the macroscopic properties of matter\cite{keimer2017physics}. Topological states of quantum matter exhibiting non-local properties emerging from its microscopic degrees of freedom have become a rapidly growing field in this regard, starting with the discovery of the first topologically non-trivial state, the quantum Hall state\cite{RevModPhys.58.519}. The quantization of the Hall conductivity in a two-dimensional electron gas (2DEG) arises due to the Berry curvature induced by the magnetic field breaking time-reversal symmetry. Classical light in the form of optical Floquet drives has recently emerged as a way to engineer the topological properties of quantum matter, exemplified by the Floquet topological insulator\cite{oka2009photovoltaic,lindner2011floquet}. Circularly polarized light in particular has shown special promise for Floquet-induced topological band structure due to its potential for also breaking time-reversal symmetry\cite{wang2013observation,mciver2020light}. Floquet engineering, where non-equilibrium light-matter states are formed by means of an external optical drive, possesses similarities to cavity quantum electrodynamics, where hybrid light-matter states are created by means of cavity strong light-matter interaction\cite{owens2022chiral, PhysRevA.97.013818, wang2009observation, PhysRevB.99.235156}. The prediction of a modified ground state containing virtual light- and matter-excitations in the ultra-strong light-matter coupling regime has raised the possibility of engineering non-trivial topologies using the quantum fluctuations of a cavity\cite{Ciuti2005}. The development of chiral cavities supporting non-degenerate circularly polarized modes by breaking time-reversal symmetry is especially important in this context. 

\begin{figure*}
	\includegraphics[width=1.0\textwidth]{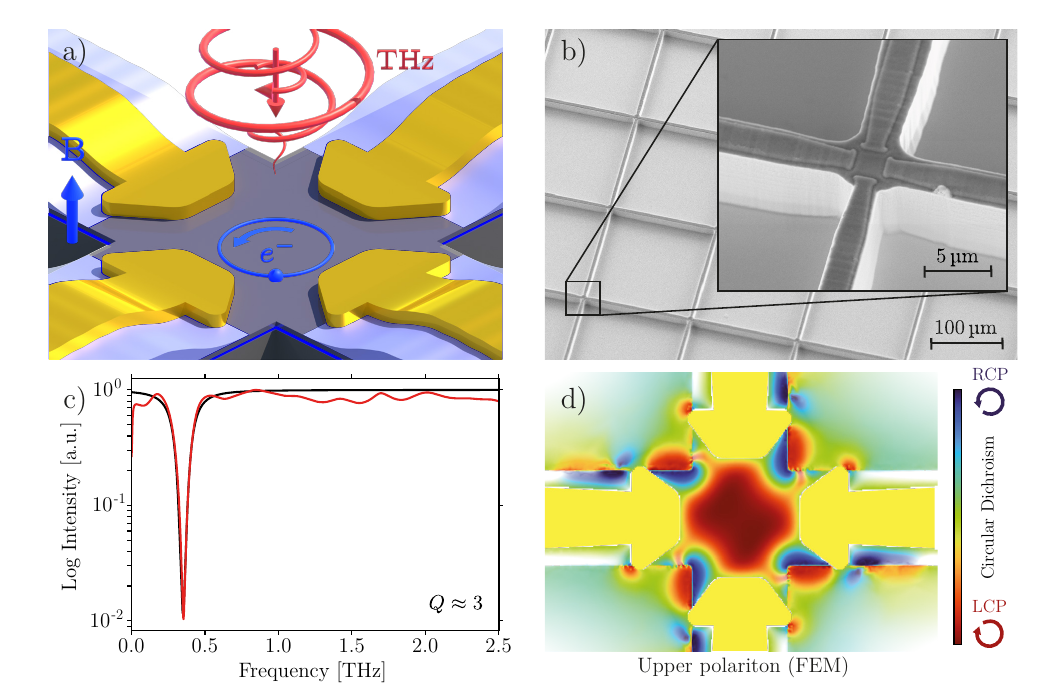}
	\caption{\label{fig:fig1}(a) The linearly polarized nano-antenna modes are coupled to the time-reversal symmetry-breaking inter-Landau level transition of the 2DEG (in blue), giving rise to non-degenerate circularly polarized modes. In a terahertz time-domain spectroscopy (THz-TDS) transmission measurement the cyclotron motion is either co-rotating or counter-rotating with the incoming circularly polarized THz pulse depending on the direction of the magnetic field, and therefore couples either to the lower and upper polariton modes or to the counter-rotating mode, respectively. (b) Scanning electron micrograph of a fabricated nano-antenna array (the inset shows a zoom-in on the unit cell). (c) Room-temperature THz-TDS transmission measurement at zero magnetic field (red curve) showing the presence of only a single well-pronounced cavity resonance together with the quality factor estimated using a Lorentzian fit (black curve). (d) Circular dichroism calculated from finite-element simulation at the upper polariton frequency with $B=B_{\mathrm{cav}} = m^*\omega_{\mathrm{cav}}/e$ illustrating the  chiral character of the mode in the region of the 2DEG. The image is weighted (using the alpha factor, i.e.\ the transparency) by the magnitude of the electric field to illustrate the degree of circular dichroism only where the field is the strongest. }
\end{figure*}

The time-reversal symmetry of a cavity can be broken by means of strong light-matter coupling to the inter-Landau level transition of a high-mobility two-dimensional electron gas in a perpendicular magnetic field\cite{hubener2021engineering}. The cyclotron motion of the electrons due to the Lorentz force has an angular frequency which lies in the terahertz (THz) range and is given by $\omega_{c}=eB/m^*$ where $e$ is the electron charge, $B$ is the magnetic flux density, and $m^*$ is the effective electron mass. At low temperatures it becomes quantized into a harmonic system of Landau levels separated in energy by $\hbar\omega_{c}$ and due to Pauli's exclusion principle and angular momentum conservation the magnetic field-dependent inter-Landau level transition consists of the two Landau levels closest to the Fermi surface $E_F$. It has a large in-plane collective dipole moment due to the large dipole moment of the electron cyclotron motion combined with the large degeneracy of the Landau levels, given by $|\vec{d}| \propto el_B\sqrt{N_e}$ with $N_{e}$ being the total electron density and $l_B = \sqrt{\hbar/eB}$ the magnetic length\cite{hagenmuller2010ultrastrong}. Therefore the rate of light-matter interaction quantified by the vacuum Rabi frequency $\Omega_R \propto \vec{d}\cdot \vec{E}_\mathrm{vac}$ can easily exceed the threshold for the ultra-strong coupling regime of $\Omega_R/\omega_\mathrm{cav}>0.1$ if this transition is coupled to the electric field $\vec{E}_\mathrm{vac}$ of a THz cavity mode, resulting in a hybrid light-matter system of Landau polaritons\cite{Scalari1323}. Metamaterial-based THz cavities containing such hybrid states possess the useful property of being easily accessible both optically using THz time-domain spectroscopy and by electronic magneto-transport measurements. In particular, since electronic measurements feature an energy scale of $\sim k_B T$, which at cryogenic temperatures is orders of magnitude smaller than the energy scale of the cavity resonance, they offer the possibility for studying the modified ground state of the ultra-strong coupling regime\cite{paravicini2019magneto,appugliese2022breakdown}. 

Fabry-Perot cavites breaking time-reversal symmetry have been demonstrated by incorporating a quantum well containing a two-dimensional electron gas\cite{mavronafabryperot}. In such cavities the two independent degenerate linearly polarized orthogonal modes are coupled by the inter-Landau level transition of the 2DEGs placed in the center. The combined system can be described in terms of a Hopfield model\cite{hopfield1958theory} consisting of two degenerate circularly polarized modes, one left-hand circularly polarized (LCP) and one right-hand circularly polarized (RCP), where only the mode co-rotating with the cyclotron resonance couples actively, whereas the other counter-rotating mode is only coupled to the cyclotron resonance via the counter-rotating terms of the minimal coupling Hamiltonian. The resulting coupled system possesses three non-degenerate circularly polarized modes\cite{kono2018}: lower and upper polariton modes co-rotating with the electron cyclotron motion and a counter-rotating mode with a dispersion referred to as the vacuum Bloch-Siegert shift, after the first-observed phenomenon in nuclear magnetic resonance under strong irradiation\cite{vbsshiftobservation}. 

An advantage of using a metamaterial cavity instead of a Fabry-Perot cavity is its deep sub-wavelength confinement, which leads to an extremely large electric field enhancement of a factor $\sim 100$ and therefore a large vacuum electric field inside the cavity $E_{\mathrm{vac}} \sim \SI{250}{\volt\per\metre}$ due to the small effective mode volume $V_{\mathrm{eff}} \sim 10^{-9}(\lambda/2n)^3$. The ultra-strong coupling regime is therefore reached with only $\sim 2000$ electrons in the gap (see Supplemental Material A\cite{appendix}). THz metamaterial cavities with a cross-like structure have previously been demonstrated\cite{Messelot:23}, exhibiting both deep sub-wavelength confinement and high quality factor. However the difficulty of obtaining two orthogonal linearly polarized modes coupled by a 2DEG in such geometries is challenging due to the significant capacitive coupling between the orthogonal directions. A half-wave dipole antenna consisting of two $\lambda/4$ dipole nano-antennas separated by a capacitor gap has a linearly polarized electric field inside the gap, and a cross-dipole or turnstile antenna consisting of two orthogonal half-wave dipole antennas with a mutual capacitor gap containing a 2DEG features the same electromagnetic components as the Fabry-Perot cavity and is a conceptually simple way to implement a chiral sub-wavelength metamaterial cavity. The chiral cavity illustrated in Fig.~\ref{fig:fig1} (a) consists of an array of nano-antennas with length $\SI{150.5}{\micro\metre}$ and width $\SI{2.0}{\micro\metre}$ and a mutual gap of $\SI{2.5}{\micro\metre}$. A scanning-electron micrograph of a fabricated array can be found in Fig.~\ref{fig:fig1} (b). An array was used in order to enhance the directionality of the antennas\cite{feuillet2013strong}. The resulting cross-dipole nano-antenna array in the absence of a magnetic field possesses a single resonance with frequency $f_{\mathrm{cav}} \approx \SI{365}{\giga\hertz}$ and a quality factor $Q\approx 3$ determined from room-temperature THz time-domain spectroscopy measurements as shown in Fig.~\ref{fig:fig1} (c). 

The antennas were defined and connected to a square mesa by a deep dry etching of \SI{10}{\micro\metre} using a \ce{SiO2} hard mask, which was chosen because of its relatively low THz refractive index of $2.1$. Since the strength of the electric field in between the orthogonal antennas is proportional to the permittivity, this reduces the stray electric field and consequently also the amount of parasitic capacitive coupling. Patches were added to further reduce the parasitic interaction between the orthogonal nano-antennas and maximize the linear electric field inside the capacitor gap that is coupled to the 2DEG. The degree of chirality of the design was quantified by the amount of \emph{circular dichroism}\cite{Poulikakos2016}
\begin{equation}\label{eq:cd}
    \mathrm{CD} = \frac{|\vec{E}_{\scriptscriptstyle\mathrm{RCP}}|^2-|\vec{E}_{\scriptscriptstyle\mathrm{LCP}}|^2}{|\vec{E}_{\scriptscriptstyle\mathrm{RCP}}|^2 + |\vec{E}_{\scriptscriptstyle\mathrm{LCP}}|^2}
\end{equation}
of the electric field in-plane with the 2DEG with values ranging from perfectly LCP, $-1$, to perfectly RCP, $+1$. Fig.~\ref{fig:fig1} (d) depicts the circular dichroism extracted from a finite element simulation of the coupled system where the 2DEG was modelled as a gyrotropic material, i.e.\ with a tensorial conductivity. The in-plane electric field of the upper polariton mode was used to illustrate that in the antenna gap the electric field is mainly LCP. Regions with differing amounts of circular dichroism can be found outside of the antenna gap, showing also that the resulting cavity is not perfectly chiral. The cavity modes of nano-antennas differ from those of Fabry-Perot cavities because the electric field is not strictly confined to the gap. The electric field from one half-wave dipole antenna will, due to residual capacitive coupling and skin effect, modify the current distribution in the orthogonal one. This in turn changes the electric field distribution also inside the gap, making the electric fields of the two orthogonal modes not be fully linear and decoupled from each other. 

Stripes of 2DEG were kept underneath the nano-antenna instead of etching patches of 2DEG in the capacitor gap as the latter would yield a 2D plasma frequency proportional to $1 / \sqrt{W}$, that with a width of $W=\SI{2.5}{\micro\metre}$ would be $\omega_p/2\pi \sim \SI{400}{\giga\hertz}$. This would lead a zero-field value of the magneto-plasmon dispersion quasi-resonant with the cavity and consequently a strong quenching of the coupling strength\cite{gian2017magnetoplasmon}. However the presence of the 2DEG below the antenna arms gives rise to a continuum of magneto-plasmon modes\cite{Cortese:19}. The deep sub-wavelength confinement of the antennas means that the wavelength is much larger than the gap, which leads to diffraction. A large fraction of the light scattered from the gap will have a significant momentum component in-plane with the 2DEGs and can be absorbed by this continuum. This leads to a reduction in transmission since these magneto-plasmon modes have large non-radiative losses due to their electric field being in-plane with the 2DEGs. Because the frequency range of this continuum of modes is just above the cavity resonance, this leads to the disappearance of the signature of the upper polariton and counter-rotating modes at low magnetic fields, an effect referred to as polaritonic non-locality\cite{rajabali2021polaritonic}. To mitigate the impact of this effect the nano-antennas are lifted off the surface in the region outside the gap by a layer of \ce{SiO2} with a thickness up to \SI{1.0}{\micro\metre}. A variation in the thickness of approximately \SI{100}{\nano\metre} was introduced into this insulating \ce{SiO2} layer by etching stripes at irregular distances and irregular widths before depositing the nano-antennas in order to further break up the continuum of magneto-plasmon modes\cite{Rajabali:23}. 

\begin{figure*}
	\includegraphics[width=1.0\textwidth]{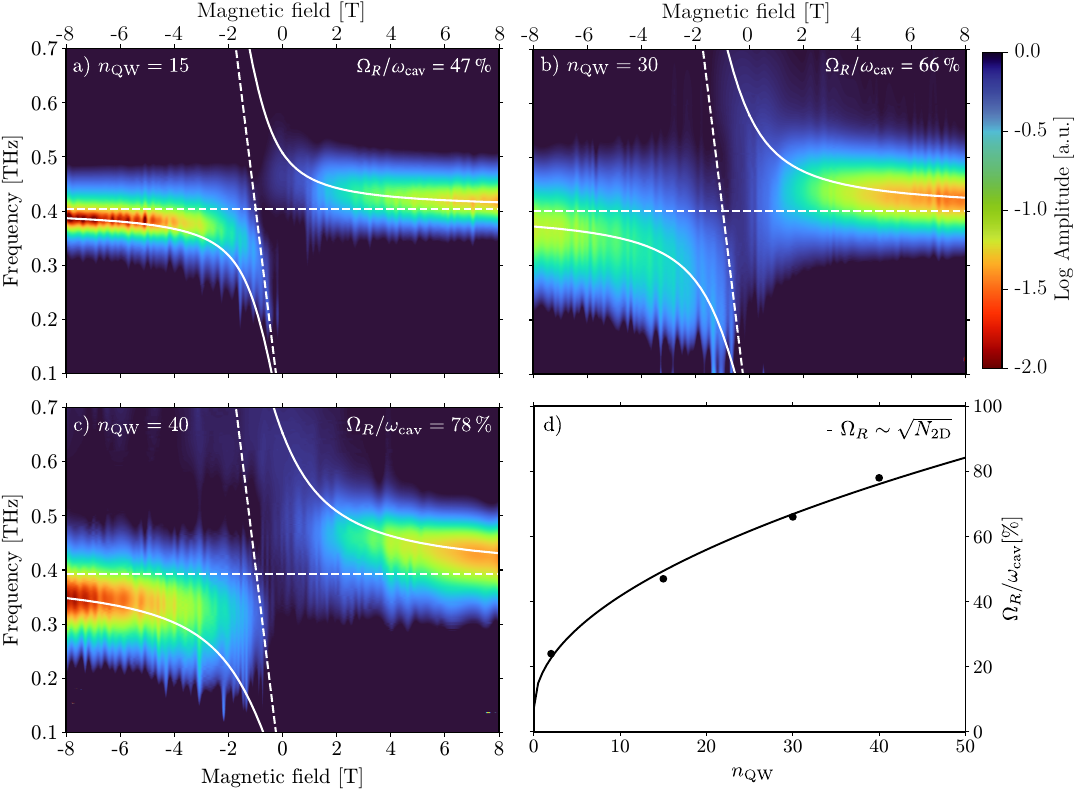}
	\caption{\label{fig:fig3}Terahertz time-domain spectroscopy measurements of left-hand circularly polarized light as a function of perpendicular magnetic field at a temperature of $T=\SI{3.2}{\kelvin}$ for samples with (a) 15 quantum wells, (b) 30 quantum wells, (c) 40 quantum wells. (d) shows the scaling of the normalized coupling strength as a function of the number of quantum wells. The excellent agreement between the normalized light-matter coupling and its expected $\sqrt{N_e}$-dependence provides good evidence for the chirality of the nano-antenna arrays.}
\end{figure*}

High-mobility single- and multi-quantum well \ce{GaAs}/\ce{AlGaAs} heterostructures were used for the fabrication of samples with different rates of light-matter interaction using standard clean-room photo-lithography techniques (see Supplemental Material D\cite{appendix}). The fabricated samples were then measured using a THz time-domain spectroscopy setup with a bandwidth of \SIrange{0.1}{3}{\tera\hertz} equipped with a split-coil superconducting magnet to probe the transmission as a function of perpendicular magnetic field from $B = -\SI{8}{\tesla}$ to $B = +\SI{8}{\tesla}$\cite{Grischkowsky:90}. Two broadband THz retarders mounted at an angle of $\SI{45}{\degree}$ were used in order to measure the transmission of circularly polarized light. A vertically polarized terahertz pulse was generated using an inter-digitated \ce{GaAs} photoconductive antenna photoexcited by the femtosecond pulses from a \SI{70}{\mega\hertz} Ti:sapphire laser, which was converted to circular polarization by the first THz retarder. The transmitted pulse after the sample was then converted back to vertical polarization by the second THz retarder for electro-optic detection, which was done using a \SI{3}{\milli\metre} $\langle 110\rangle$ \ce{ZnTe} crystal. The incoming circularly polarized THz pulse is co-rotating with the electron cyclotron motion for $B < \SI{0}{\tesla}$ and therefore only couples to the co-rotating lower and upper polariton modes in this region whereas for $B > \SI{0}{\tesla}$ it couples to the counter-rotating mode. The cavities exhibit excellent agreement with the dispersion of a single-mode Hopfield model\cite{hagenmuller2010ultrastrong} when probed with linear polarization (see Fig. 7 in Supplemental Material) and the degree of chirality can therefore be determined by the degree of asymmetry between positive and negative magnetic fields and the agreement with the dispersion of the Hopfield model with circularly polarized modes (see Supplemental Material A\cite{appendix} and references \cite{kono2018,hopfield1958theory, hagenmuller2010ultrastrong} therein). Measurements of samples with different number of quantum wells were used to study the system as a function of coupling strength (see Fig.~\ref{fig:fig3}). 

Despite the measures taken the lower polariton and counter-rotating modes still exhibit significant broadening due to polaritonic non-locality, which increases with the number of quantum wells. The upper polariton branch is also barely visible, as can be seen in Fig.~\ref{fig:fig3}. The lower polariton mode in the multi-quantum well samples also shows significant deviations from the Hopfield model, implying that despite the stripe geometry there is still the presence of a magneto-plasmon dispersion. A consequence of the mesa design is that the strip geometry significantly modifies the Coulomb interaction configuration within the system. This can modify the dispersion of the magneto-plasmon modes and introduce a new category of low-frequency edge magneto-plasmon modes that are independent of the antenna\cite{xia1994edge}. Since these modes exist purely because of the stripe geometry they are not mitigated by lifting the antenna off the surface, and although they possess an acoustic dispersion and would consequently normally not be excited, because of the deep sub-wavelength confinement and subsequent diffraction they could induce significant absorption, and therefore also contribute to the disappearance of the polariton dispersion at low magnetic fields, in addition to modifying the dispersion of the lower polariton mode. To compensate for these effects the coupling strength was fitted only with the lower polariton mode at $B=\SI{-8}{\tesla}$ and the counter-rotating mode at $B=+\SI{8}{\tesla}$. The normalized coupling strengths $\Omega_R/\omega_{\mathrm{cav}}$ are however in excellent agreement with the expected $\sqrt{N_e}$-dependence of the total electron density, providing strong evidence for the chiral nature of the metamaterial. 

\begin{figure*}
	\includegraphics[width=1.0\textwidth]{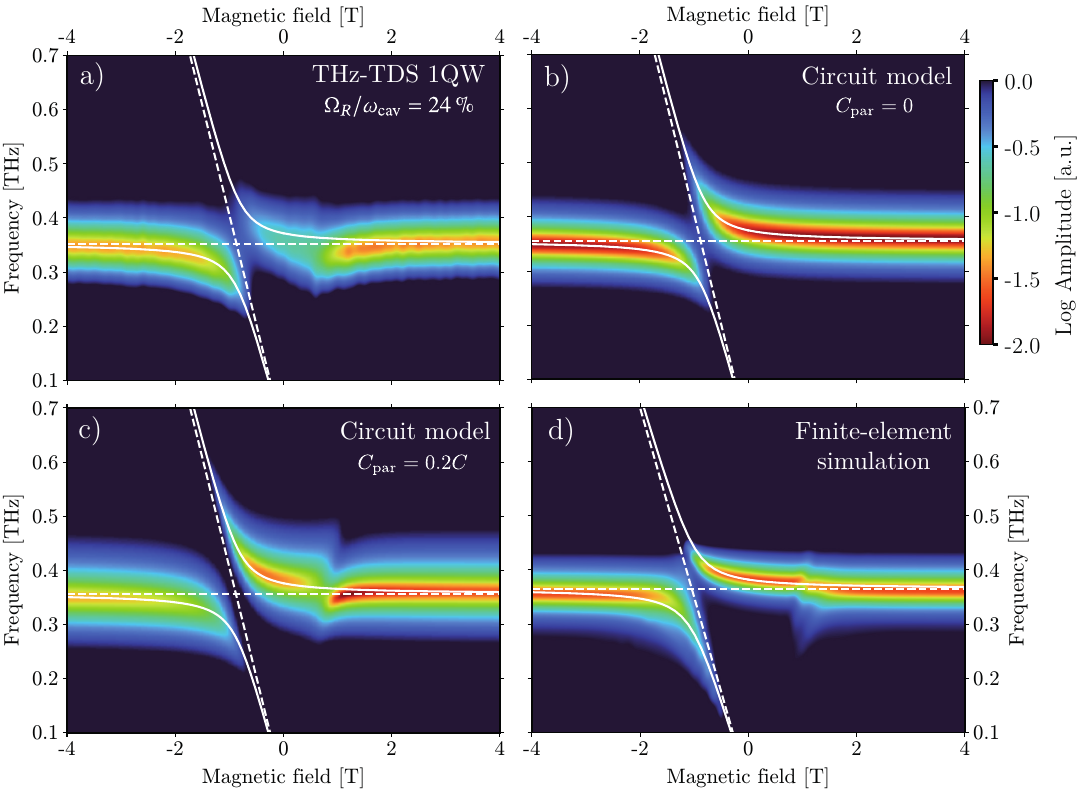}
	\caption{\label{fig:fig2}(a) THz-TDS measurement for a single quantum well sample as a function of perpendicular magnetic field together with a fit of the Hopfield model with circular polarization\cite{kono2018} using the LP mode. (b) and (c) shows the simulated transmission of the circuit model with and without parasitic capacitive coupling between the orthogonal antennas, respectively. (d) Finite-element simulation of the transmission of the cavity, modelling the 2DEG as a gyrotropic material. }
\end{figure*}

The measurement of a single quantum well sample can be found in Fig.~\ref{fig:fig2} (a) together with the fitted normalized coupling strength. The dispersion shows excellent agreement for the LP mode, suggesting that in contrast to the multi-quantum well samples there is no magneto-plasmon dispersion. The counter-rotating mode however is suppressed and deviates in the region of the cavity resonance. The vanishing of the counter-rotating and upper polariton modes at low magnetic fields can be understood in terms of the polaritonic non-locality, which is still present despite the measures taken to mitigate it. To understand the deviation it should be noted that the deep sub-wavelength confinement of THz nano-antennas renders them suitable for modelling as lumped-element RLC circuits coupled by a gyrotropic material in the capacitor gap. It can be shown (see Supplemental Material B\cite{appendix}, including references \cite{ihn2009semiconductor,balanis1996antenna,kraus1950antennas, cyclotrondecaytime}) that a circuit model that takes the broken time-reversal symmetry into account possesses the same dispersion as the Hopfield model for a chiral Fabry-Perot cavity, provided that there is negligible capacitive coupling between the nano-antennas outside the 2DEG region. In Fig.~\ref{fig:fig2} (b) the circuit transmission model in the absence of parasitic interaction between the antennas is shown, together with the Hopfield dispersion. When a parasitic interaction is added, the counter-rotating mode at magnetic fields $B<B_{\mathrm{cav}} \equiv m^*\omega_{\mathrm{cav}}/e$ competes with the LP mode due to the parasitic interaction in the case of a single quantum well, as can be seen in Fig.~\ref{fig:fig2} (c). The effects of the parasitic interaction include a modified dispersion and loss of the modes, an opening of a gap in the dispersion of the counter-rotating mode at $\omega_\mathrm{cav} = \omega_{c}$ and magnetic field-dependent polarization states of the cavity modes. This model also explains why the effect of the parasitic interaction decreases with electron density and is therefore smaller in the multi-quantum well samples. Minimizing its effect requires minimizing the stray electric field outside the gap and maximizing the intra-cavity one. For such field enhancement long nano-antennas were chosen to make the inductance as large as possible with respect to the capacitor gap for a given cavity frequency $\omega_{\mathrm{cav}} \approx 1 / \sqrt{L(C-C_{\mathrm{par}})}$ where $C_{\mathrm{par}}$ denotes the parasitic capacitance between the orthogonal antennas.

\section*{Conclusion and Outlook}

We have demonstrated a THz chiral sub-wavelength cavity on the basis of including a time-reversal symmetry breaking element, a 2DEG in a perpendicular magnetic field. The discrepancies between measurements and the circuit and Hopfield models can be understood within the framework of parasitic capacitive coupling between the orthogonal nano-antennas and polaritonic non-locality due to their deep sub-wavelength confinement. The issue of polaritonic non-locality remains significant despite the measures taken to mitigate it, with the measures only being effective in the single quantum well sample. This is most likely because the distance between the nano-antennas and the 2DEGs is still smaller than the total thickness of the multi-quantum well heterostructure, even for $n_{\mathrm{QW}} = 15$. However its effects of reducing transmission at low magnetic fields and vanishing of the upper polariton mode does not impact the degree of chirality of the cavity. In a next step the direct effect of the electromagnetic vacuum field on the 2DEG can be investigated by means of magneto-transport measurements in a van der Pauw configuration\cite{philips1958method}, taking advantage of the symmetry of the chiral cavity. 

By further reducing the gap size this system also offers the possibility of exploring the few-electron regime beyond the dilute regime of few excitations in which the Hopfield model is valid\cite{kellerfewelectron}. Additionally it can be used with a variety of 2D materials such as graphene, instead of the \ce{GaAs}/\ce{AlGaAs} quantum wells used here. 

Since emphasis was placed on conceptual simplicity in the design other geometries that would suffer less from parasitic capacitive coupling between the modes can be explored, for example ones where the gap is surrounded by three antennas instead of four. 

The effect of the circularly polarized ground state of the cavity on other systems could be studied by bringing them into close physical proximity with its capacitor gap. However one aspect, that has not been fully explored in this work, are the effects of using an array of nano-antennas compared to just a single cross, in particular how the chiral character of the electric field inside one gap is modified by its neighbors since they are coupled via their mutual inductance. If the array can be utilized as a coupled cavity system to impose a chiral electric field distribution onto a gap without 2DEG this could enable studying the effect of a chiral electromagnetic ground state on other systems, with the caveat of the presence of the external magnetic field. 

\begin{acknowledgements}
The authors acknowledge the FIRST-Lab cleanroom where the majority of the fabrication process took place, S. Rajabali for her work on the nano-antennas that served as a starting point for the design of the chiral metamaterial cavities and her ideas on mitigating the issue of polaritonic non-locality, A. Rubio for discussions, and the funding and support provided by the Swiss National Science Foundation (SNF) through the National Centre of Competence in Research Quantum Science and Technology (NCCR QSIT) and by the Alexander von Humboldt Stiftung through the Humboldt Research Award awarded to J. Faist.
\end{acknowledgements}

\clearpage
\appendix
%\maketitle
\begin{center}
	\textbf{Supplemental Material: Terahertz chiral sub-wavelength cavities breaking time-reversal symmetry via ultra-strong light-matter interaction}
\end{center}
\section{Quantum mechanical theory}\label{appendix:hopfield}

\begin{figure}
	\includegraphics[width=0.8\columnwidth]{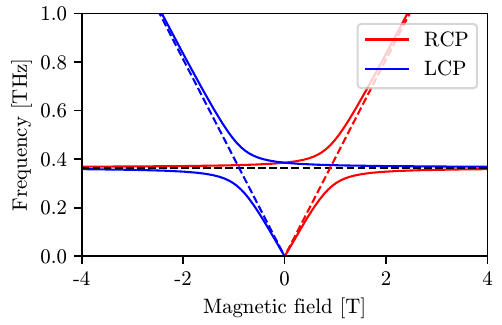}
	\caption{\label{fig:hopfielddisppoly}Eigen-frequencies for RCP and LCP obtained by solving Eq.~\ref{eq:rcpdisp} and Eq.~\ref{eq:lcpdisp} for a cavity with $\omega_\mathrm{cav}/2\pi=\SI{365}{\giga\hertz}$ and normalized coupling strength $\bar{\Omega}_R = 0.24$.}
\end{figure}

Following Ref.~\cite{kono2018}, we describe the interaction between the 2DEG-embedded metamaterial cavity in a static perpendicular magnetic field with circularly polarized light within an Hopfield-Bogoliubov Hamiltonian model. The second-quantized Hamiltonian, according to Ref.~\cite{kono2018}, is given by
\begin{equation}\label{eq:HopBogHam}
	\mathcal{H} = \mathcal{H}_\mathrm{cav} + \mathcal{H}_\mathrm{Landau} + \mathcal{H}_\mathrm{int} + \mathcal{H}_\mathrm{dia}
\end{equation}
where 
\begin{equation}
	\mathcal{H}_\mathrm{cav} = \sum_{\xi=\pm}\sum_{n_z} \hbar\omega^{n_z}_\mathrm{cav}\left(a^\dagger_{n_z,\xi}a_{n_z,\xi} + \frac{1}{2} \right)  
\end{equation}
is the cavity Hamiltonian, $\xi$ labels the right ($+$) and left ($-$) circular polarization components, $n_z$ labels the cavity modes, and $a_{n_z,\xi}^\dagger$ and $a_{n_z,\xi}$ are the creation and annihilation operators for the mode $n_z$ with polarization~$\xi$;
\begin{equation}
	\mathcal{H}_\mathrm{Landau} = \hbar\omega_c \left(b^\dagger b + \frac{1}{2} \right)
\end{equation}
is the 2DEG Hamiltonian, expressed in terms of $b^\dagger$ and $b$, which are creation and annihilation operators of the bright collective excitation of cyclotron motion (CR) with angular frequency $\omega_c$, i.e.\ the inter-Landau level transition described in the main text. In the diluted limit, in which the number of excited electrons is far less than the total number of electrons in the 2DEG, $\comm{b}{b^\dagger} \simeq 1$;
\begin{equation}
	\mathcal{H}_\mathrm{int} = \sum_{n_z} i\hbar \Omega_{R}^{n_z} \left[b^\dagger (a_{n_z,+} + a^\dagger_{n_z,-}) - b(a_{n_z,-} + a^\dagger_{n_z,+}) \right]
\end{equation}
is the interaction Hamiltonian, which does consider also the counter-rotating terms. Notice that $b$ and $b^\dagger$ are coupled to creation and annihilation operators having opposite circular polarizations, in particular the cyclotron resonance interacts with right circularly polarized photons in the co-rotating manner, and with the left circularly polarized photons in the counter-rotating manner. The $i$ coefficient follows from a phase choice, but a phase choice in which the interaction Hamiltonian is real is also possible. In Ref.~\cite{kono2018} the interaction rate at the cyclotron frequency for mode $n_z$, that is $\Omega_{R}^{n_z}$, is calculated for the case in which the cavity is a 1D THz photonic-crystal cavity. In our case we will consider $\Omega_{R}^{n_z}$ as an independent parameter, and obtain its value via a fitting procedure;
\begin{equation}
	\mathcal{H}_\mathrm{dia} = \sum_{n_z, n_z'} \frac{\hbar \Omega_{R}^{n_z}\Omega_{R}^{n_z'}}{\omega_c} (a_{n_z,-} + a^\dagger_{n_z,+})(a_{n_z',+} + a^\dagger_{n_z',-})
\end{equation}
is the diamagnetic Hamiltonian, which arises from the square modulus of the vector potential. Notice that it is a second order term in which pairs of cavity photons are involved, belonging to the same mode (\emph{self-energy} terms) or to different modes (\emph{cross} terms). The Hamiltonian~(\ref{eq:HopBogHam}) is diagonalized following the  procedure of Ref.~\cite{hopfield1958theory} for the case of a single cavity mode, i.e. $n_z = 1$ (and therefore $\Omega_R^{n_z} \equiv \Omega_R$ and $\omega_{\mathrm{cav}}^{n_z} \equiv \omega_{\mathrm{cav}}$) by defining the polariton annihilation operators
\[p_{j,k} = w_{j,k}^{+}\hat{a}_{+} + w_{j,k}^{-}\hat{a}_{-} + x_{j,k}\hat{b} + y_{j,k}^{+}\hat{a}_{+}^\dagger + y_{j,k}^{-}\hat{a}_{-}^\dagger +  z_{j,k}\hat{b}^\dagger\]
and computing $[\hat{p}_{j,k}, \mathcal{H}]$. This gives the Hopfield matrix
\begin{widetext}
	\[M = \begin{pmatrix}\omega_{\mathrm{cav}}+D & 0 & -i\Omega_R & 0 & D & 0 \\ 0 & \omega_{\mathrm{cav}}+D & 0 & D & 0 & i\Omega_R \\ i\Omega_R & 0 & \omega_{c} & 0 & i\Omega_R & 0 \\ 0 & -D & 0& -\omega_{\mathrm{cav}}-D & 0 & -i\Omega_R \\ -D & 0 & i\Omega_R & 0 & -\omega_{\mathrm{cav}}-D & 0 \\ 0 & i\Omega_R & 0 & i\Omega_R & 0 & -\omega_{c}\end{pmatrix}\]
\end{widetext}
in the basis of the Hopfield coefficients $\vec{p}_{j,k} = (w_{j,k}^{+}, w_{j,k}^{-}, x_{j,k}, y_{j,k}^{+}, y_{j,k}^{-}, z_{j,k})$. Note that we have introduced the diamagnetic factor $D\equiv \Omega_R^2/\omega_{c}$. This matrix satisfies $M\vec{p}_{j,k} = \omega_{j,k}\vec{p}_{j,k}$ and by introducing the magnetic field-dependent coupling
\begin{equation}
	\Omega_{R} = \bar{\Omega}_R\sqrt{\omega_\mathrm{cav} \omega_c},   
\end{equation}
where $\bar{\Omega}_R = \Omega_R/\omega_{\mathrm{cav}}$ is the normalized coupling for $\omega_{\mathrm{cav}} = \omega_{c}$ we obtain that the eigen-frequencies $\omega$ satisfy the equations
\begin{equation}\label{eq:rcpdisp}
	\omega^3 - \omega_{c}\omega^2 - \omega_{\mathrm{cav}}^2(1 + 2\bar{\Omega}_R^2)\omega + \omega_{\mathrm{cav}}^2\omega_{c} = 0 % This should indeed contain the normalized coupling (I replaced OmegaR with \bar{\Omega}_R sqrt(omega_c omega_cav) in the calculation
\end{equation}
\begin{equation}\label{eq:lcpdisp}
	\omega^3 + \omega_{c}\omega^2 - \omega_{\mathrm{cav}}^2(1 + 2\bar{\Omega}_R^2)\omega - \omega_{\mathrm{cav}}^2\omega_{c} = 0
\end{equation}
A careful analysis of the eigen-vectors, which amounts to looking at the square modulus of the coefficients of the $a_{\pm}$ and $a^\dagger_{\pm}$ components for both the cavity modes (see Figure~\ref{fig:hopfielddisppoly}), allows us to infer that Eq.~\ref{eq:rcpdisp} describes the modes coupled with right circularly polarized light ($+$) and Eq.~\ref{eq:lcpdisp} the modes coupled with left circularly polarized light ($-$). Notice that by reversing the magnetic field we get ($\omega_c \rightarrow -\omega_c$), so that the the equations with upper and lower signs will describe the interaction with left and right circularly polarized light, respectively (the time-reversal symmetry is indeed preserved by the Hamiltonian). Moreover, we see that the equations stay the same under the substitutions $\omega \rightarrow -\omega$ and $\omega_c \rightarrow -\omega_c$, so the solutions must be symmetric about the origin of the magnetic field-frequency plane.

\subsection{Fitting procedure}

The multiple quantum well samples were fitted by only considering the difference between the VBS shift at $B=+\SI{8}{\tesla}$ and the LP mode at $B=\SI{-8}{\tesla}$. The VBS and LP solutions to Eq.~\ref{eq:rcpdisp} are given by
\begin{widetext}
	\begin{align}\label{eq:hamsol}
		\omega_{\mathrm{VBS}} &= -\frac{\omega_{c}}{3} + \frac{2}{3}\sqrt{\omega_{c}^2 + 3\omega_{\mathrm{cav}}^2(1+2\bar{\Omega}_R)}\cos\left(\frac{1}{3}\arg(C)\right) \\
		\omega_{\mathrm{LP}} &= \frac{1}{3}\left[\omega_{c} + \sqrt{\omega_{c}^2 + 3\omega_{\mathrm{cav}}^2(1+2\bar{\Omega}_R)}\left(\cos\left(\frac{1}{3}\arg(C)\right) - \sqrt{3}\sin\left(\frac{1}{3}\arg(C)\right)\right)\right] \\
		C &\equiv -\omega_{c}^3-9\omega_{\mathrm{cav}}^2\omega_{c}(\bar{\Omega}_R^2-1)+3i\sqrt{3}\omega_{\mathrm{cav}}\sqrt{\omega_{c}^4+\omega_{\mathrm{cav}}^4(2\bar{\Omega}_R^2+1)^3 + \omega_{\mathrm{cav}}^2\omega_{c}^2(\bar{\Omega}_R^4 + 10\bar{\Omega}_R^2 - 2)}
	\end{align}
\end{widetext}
and the fitting was done using the least-squares method with respect to the variables $\omega_\mathrm{cav}$ and $\bar{\Omega}_R$. For the single quantum well sample only the LP mode was used because the frequency difference between the LP and VBS modes was below the approx.~\SI{40}{\giga\hertz} resolution of the THz-TDS system employed and because the VBS shift was significantly modified by the parasitic capacitive interaction between the antennas. 

\subsection{Coupling strength}

The vacuum Rabi frequency is given by
\begin{equation}
	\Omega_R = \frac{1}{\hbar}\vec{d}\cdot\vec{E}_\mathrm{vac}\sqrt{N_e/\nu}
\end{equation}
where $|\vec{d}| = el_B\sqrt{\nu}$ is the dipole moment of a single electron and $\nu = hn_{2D}/(eB)$ is the filling factor corresponding to the number of filled Landau levels, $n_{2D}$ being the sheet electron density. The vacuum electric field is given by
\begin{equation}
	|\vec{E}_\mathrm{vac}| = \sqrt{\frac{\hbar\omega_\mathrm{cav}}{\varepsilon_r\varepsilon_0 V_{\mathrm{eff}}}}
\end{equation}
where $V_{\mathrm{eff}}$ denotes the effective mode volume. The number of coupled electrons in the gap is given by the population of the top-most filled Landau level below the Fermi energy and since the total number of electrons in the gap is $N_e = n_{QW}n_{2D}L_xL_y$ this number can be estimated as $N_e/\nu$. With a gap of \SI{2.5}{\micro\metre} and a single quantum well with $n_{2D}=\SI{4.5e11}{cm^{-2}}$ this means that at the cavity resonance frequency $f_{cav}=\SI{365}{\giga\hertz}$, which corresponds to a magnetic field of $B_{cav}\approx \SI{0.9}{\tesla}$, the number of electrons is
\begin{equation}
	N_e/\nu \approx 2000
\end{equation}

\subsection{Magneto-plasmon dispersion}

The 2D plasma frequency for a stripe of width $W$ is given by
\begin{equation}
	\omega_p = \sqrt{\frac{n_{2D}e^2\pi}{2m^*\varepsilon_r\varepsilon_0 W}}
\end{equation}
and this can in principle give rise to a magneto-plasmon dispersion
\begin{equation}
	\omega_{MP} = \sqrt{\omega_{c}^2 + \omega_p^2}.
\end{equation}
For the purposes of fitting however this was omitted due to the absence of a well-defined length because of the stripes. 

\section{Circuit Theory}\label{appendix:circuits}

\begin{figure*}
	\includegraphics[width=0.99\textwidth]{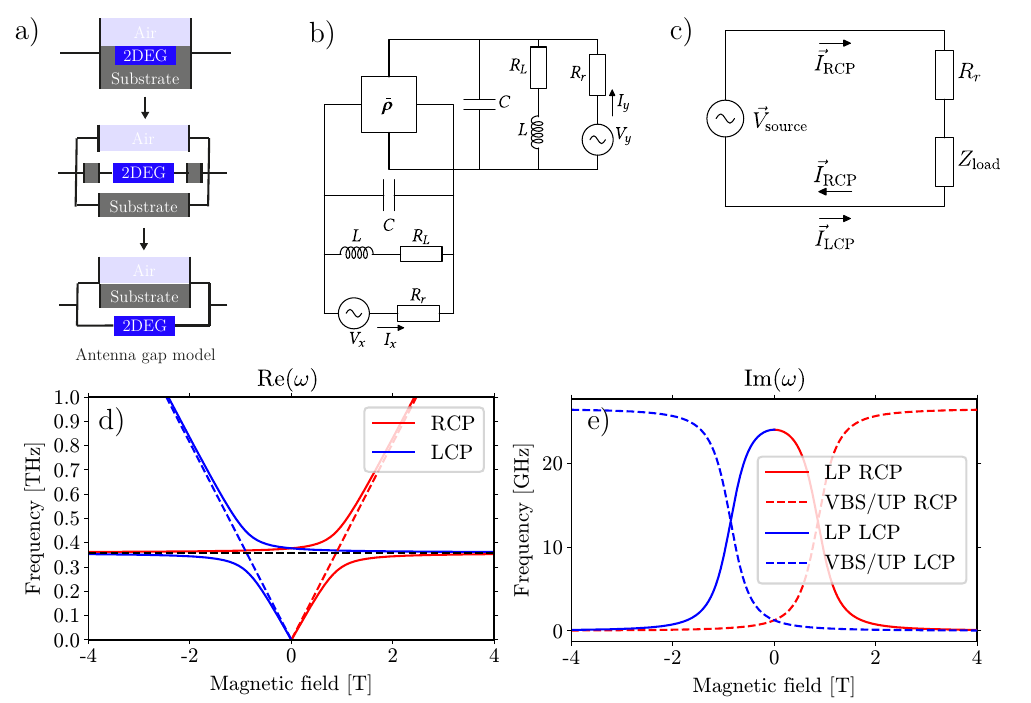}
	\caption{\label{fig:circuit_polaritons} (a) The antenna gap viewed as a capacitor with the 2DEG inside. The two-dimensional electron gas (2DEG) is capacitively coupled to the antenna by two capacitors with a capacitance far larger than the capacitance of the antenna itself, consequently their impedance is negligible and the antenna capacitor can be modelled as being in parallel with the 2DEG. (b) Circuit model of the system. The 2DEG resistivity $\vb*{\bar{\rho}}$ couples the two LC circuits which represent the arms of the $x$ and $y$ oriented antennas, which are being excited by an electric voltage with orthogonal components $V_x$ and $V_y$ (their phase difference is $\pi/2$ for circularly polarized light). (c) Model for forward- and backward-propagating electromagnetic radiation. (d) Real part of the solution for the circuit model. (e) Imaginary part of the solution for the circuit model.}
\end{figure*}

In Figure~\ref{fig:circuit_polaritons} we display the circuit model of the system: the 2DEG is represented by a four-terminal element, whose tensorial conductivity $\vb*{\bar{\sigma}}$ is given by
\begin{equation}
	\vb*{\bar{\sigma}} = \pmqty{\sigma_{xx} & -\sigma_{xy} \\ \sigma_{xy} & \sigma_{xx}}
\end{equation}
where
\begin{align}
	&\sigma_{xx} = \sigma_\mathrm{\footnotesize{DC}} \frac{1+i\omega\tau}{(1+i\omega\tau)^2 + (\omega_c\tau)^2} \\
	&\sigma_{xy} = \sigma_\mathrm{\footnotesize{DC}} \frac{\omega_c\tau}{(1+i\omega\tau)^2 + (\omega_c\tau)^2} \\
	&\sigma_\mathrm{\footnotesize{DC}} = \frac{n_{2D} e^2 \tau}{m_\mathrm{{eff}}}.
\end{align}
The resistivity $\vb*{\bar{\rho}}$ is just given by $\vb*{\bar{\rho}} = \vb*{\bar{\sigma}}^{-1}$.
The tensorial conductivity has been obtained by solving Boltzmann's transport equation~\cite{ihn2009semiconductor} for the case of free electrons confined on a two-dimensional plane, subject to a static magnetic field $B$ applied orthogonal to the plane, and driven by an oscillating in-plane electric field $E(t) = E_\omega e^{i\omega t}$, which is the field of the cavity. The surface density of electrons is $n_{2D}$, and their effective mass is $m_\mathrm{eff}$, which takes into account the effect of the lattice in which the electrons are confined. The static magnetic field forces the electron motion on circular orbits (in the absence of the driving electric field), with angular frequency $\omega_c = \abs{e}B/m_\mathrm{eff}$ (i.e.\ the cyclotron frequency), which competes with the frequency $\Gamma_c = 1/\tau$ with which the electrons are scattered by crystal defects, charged impurities, or phonons, where $\tau$ is the mean scattering time. Furthermore, since the (intensive) conductivity is in \SI{}{\ohm^{-1}}, we get the (extensive) conductance by multiplying the former by the side $l$ of the (square) 2DEG region.

The 2DEG effectively couples the pair of cross antennas (see Fig.~\ref{fig:circuit_polaritons}a), which are represented by equivalent parallel LC circuits (for the antennas oriented along the $x$ and the $y$ axis), whose bare resonance frequency is $\omega_\mathrm{cav} = 1/\sqrt{LC}$. The resistance of the antenna is given by both $R_L$ (in series with the inductor) and $R_r$ (in series with the voltage source), where $R_L$ represents the losses due to heating (i.e.\ the Joule effect), and $R_r$ is the \emph{radiation resistance}, that accounts for the power taken away from the circuit by the electromagnetic wave which is scattered by the antenna~\cite{balanis1996antenna}, and can be thought as a source resistance~\cite{kraus1950antennas}. We can define the cavity linewidth $\Gamma_\mathrm{cav} = 1/(\sigma_\mathrm{\footnotesize{DC}} R_L)$, which competes with the cavity frequency $\omega_\mathrm{cav}$, in the same way as the 2DEG linewidth $\Gamma_c$ competes with the cyclotron frequency $\omega_c$. The parallel configuration of the circuit is chosen because the electric field of the electromagnetic radiation which impinges on the sample (i.e.\ the THz beam in the TDS setup) is applied simultaneously across the antenna arm, the antenna gap, and the 2DEG region. The AC voltage sources $V_x$ and $V_y$ model the two orthogonal polarizations (along the $x$ or the $y$ axis) of the impinging radiation, and they are out of phase by $\pi/2$ when circularly polarized light is employed.

The total impedance of each antenna LC circuit is given by
\begin{equation}
	Z = \frac{1}{i\omega C + 1 / (i\omega L + R_L)},
\end{equation}
and Kirchhoff's Law states that
\begin{equation}\label{eq:Kirchhoff}
	\pmqty{V_x \\ V_y} -R_r\pmqty{I_x \\ I_y} - \vb*{\bar{G}}^{-1} \pmqty{I_x \\ I_y} = 0,
\end{equation}
where the total conductance $\vb*{\bar{G}}$ is given by
\begin{equation}
	\vb*{\bar{G}} = l \vb*{\bar{\sigma}} + \frac{1}{Z} \mathbb{I}_2,
\end{equation}
and $\mathbb{I}_2$ is the $2\times2$ identity matrix.

To find the normal modes of the coupled circuits one has to solve Eq.~\ref{eq:Kirchhoff} in the absence of the external excitation (which does not mean to set $V_x = V_y = 0$, but to entirely \emph{remove} from the circuit both the source and the radiation resistance, thus setting $I_x = I_y = 0$)
\begin{equation}
	\begin{gathered}
		\vb*{\bar{G}}\pmqty{V_x \\ V_y} = 0 \\
		\Rightarrow\quad \begin{pmatrix}
			\sigma_{xx} + 1/Z & -\sigma_{xy} \\
			\sigma_{xy} & \sigma_{xx} + 1/Z
		\end{pmatrix} \pmqty{V_x \\ V_y} = 0,
	\end{gathered}
\end{equation}
which has a nontrivial solution only if 
\begin{equation}
	\det\left(\vb*{\bar{G}}\right) = 0.
\end{equation}
By expanding the previous equation one gets

\begin{widetext}
	\begin{equation}\label{eq:polaritons_circuit}
		\omega^3 + \left[\mp\omega_c -i \left(\Gamma_c + \frac{\Gamma_\mathrm{cav}\omega_\mathrm{cav}^2}{2g^2} \right)\right] \omega^2
		- \left( 2g^2 + \omega_\mathrm{cav}^2 + \frac{\Gamma_\mathrm{cav}\Gamma_c \omega_\mathrm{cav}^2}{2g^2}
		\pm i\frac{\Gamma_\mathrm{cav} \omega_\mathrm{cav}^2\omega_c}{2g^2} \right)\omega +\left(\pm \omega_c +i \big(\Gamma_\mathrm{cav} + \Gamma_c \right)\big)\omega_\mathrm{cav}^2  = 0
	\end{equation}
\end{widetext}

where we have defined $g = \sqrt{\sigma_\mathrm{\footnotesize{DC}} / 2C\tau}$, and which has exactly the same structure as the one obtained via the Hamiltonian formalism: indeed for $\Gamma_c = \Gamma_\mathrm{cav} = 0$ (that is in the case of no losses) one obtains Eqs.~\ref{eq:rcpdisp} and~\ref{eq:lcpdisp}, and thus we can justify our definition of $g$ as the magnetic field-independent coupling between the cyclotron resonance and the cavity mode, i.e.\ $g = \bar{\Omega}_{R}$. It should be noted that due to both the antenna and the 2DEG losses, the new modes which arise from the hybridization of the cavity mode with the cyclotron resonance are not truly normal modes, as it is evident from the non-zero imaginary part of their frequencies. Notice that the circuital approach allows us to predict that the coupling will decrease as $1/\sqrt{C}$ when the capacitance of the gap between the two antennas is increased, and also that the coupling will increase with the carrier density as $\sqrt{n_s}$, in the same way as it is found in Ref.~\cite{hagenmuller2010ultrastrong}. 

In Figure~\ref{fig:circuit_polaritons} the real (c) and imaginary (d) part of the solutions of Eq.~\ref{eq:polaritons_circuit} are displayed, for the interaction of both right (upper signs in the equation) and left (lower signs) circularly polarized light. For high positive (negative) values of the magnetic field $B$ the imaginary part of the lower polariton and of the VBS mode frequencies goes to zero, indicating that the solution is a normal mode of the system. On the other hand, the nonzero imaginary part indicates a lossy mode, that can be thought only as an out-of-equilibrium (driven by the impinging radiation) hybridization of the cavity mode with the cyclotron resonance.

The circuital approach allows us to derive in a straightforward way the transmission of the system as a function of the static magnetic field and of the frequency of the impinging radiation. As it was already discussed above, the AC voltage sources $V_x$ and $V_y$ in the circuit of Fig.~\ref{fig:circuit_polaritons}~(b) provide an excitation to the circuit which is directly proportional to the intensity of the electric field components along $x$ and $y$ of the impinging electromagnetic radiation. Following Ref.~\cite{balanis1996antenna}, the electromagnetic wave which is scattered by the antennas (coupled with the 2DEG) will have electric field components directly proportional to the current which goes through the radiation resistance
\begin{equation}
	E_{x,y} \propto R_r\, I_{x,y}.
\end{equation}
To compare the results of the circuital approach with the actual transmission spectra obtained with THz time-domain spectroscopy we can calculate the power transmitted by the antenna as
\begin{equation}\label{eq:circuit_trasm}
	P = \frac{1}{2} R_r (\abs{I_x}^2 + \abs{I_y}^2),
\end{equation}
where we are summing the current contribution on both axis in the case in which we collect both components of light polarization. In Fig. 3 (b) in the main text we show a colour-plot of the logarithm of the transmission obtained with Eq.~\ref{eq:circuit_trasm}, in the case in which $V_x$ and $V_y$ have the same amplitude but are out of phase by $\pi/2$, that is the impinging radiation is circularly polarized. Notice how the polariton frequency at high values of the static magnetic field is different between $B>0$ and $B<0$. This \emph{polaritonic gap} will be closed only for much higher values of $B$. Moreover, notice how the polariton signature quickly fades out moving away from the cavity frequency, due to the nonzero value of the imaginary part of the eigenvalues.

\subsection{Effect of broken time-reversal symmetry}

Before we can discuss the effect of parasitic capacitance on the system we have to consider the effect of broken time-reversal symmetry. Because we are interested in the response of circularly polarized light we will consider RCP and LCP polarizations only. A voltage source representing an incoming electromagnetic field can be interpreted as either a forward- or backward-propagating wave. To avoid having to consider the propagation direction, we note that a backwards-propagating RCP is forwards-propagating -LCP seeing a magnetic field $-B$ if we switch the perspective between source and observer, which corresponds to a transformation $x\to -x$. Therefore we can consider two forward-propagating waves, one RCP and one -LCP. These are connected by the fact that although the antenna radiates the same polarization, it will dissipate voltage of the opposite. If we consider incoming RCP $\vec{V}_R = \frac{V}{\sqrt{2}}\begin{pmatrix}1 \\ -i\end{pmatrix}$ and -LCP $\vec{V}_L = \frac{V}{\sqrt{2}}\begin{pmatrix}-1 \\ -i\end{pmatrix}$ the radiated power is given by Eq.~\ref{eq:trscircuit}
\begin{widetext}
	\begin{equation}\label{eq:trscircuit}
		\begin{pmatrix}
			\vec{V}_{R}^{rad} \\
			\vec{V}_{L}^{rad}
		\end{pmatrix}=
		\begin{pmatrix}
			\underline{R}_r\underline{Z}(B)^{-1} & \mathbb{I}_2-\underline{R}_r\underline{Z}(-B)^{-1} \\
			\mathbb{I}_2-\underline{R}_r\underline{Z}(B)^{-1} & \underline{R}_r\underline{Z}(-B)^{-1}
		\end{pmatrix}
		\begin{pmatrix}
			\vec{V}_{R} \\
			\vec{V}_{L}
		\end{pmatrix}
	\end{equation}
\end{widetext}
If $\vec{V}_R$ is an eigenmode of the circuit, i.e. $Z(B)^{-1}\vec{V}_R = \lambda\vec{V}_R$ and $Z(-B)^{-1}\vec{V}_L = \lambda\vec{V}_L$ for some eigenvalue $\lambda$ then the off-diagonal terms disappear. Because $\vec{V}_L^{rad}$ represents the backwards-propagating RCP wave, it can be interpreted as the reflected electromagnetic radiation. 

\subsection{Parasitic coupling}

\begin{figure*}
	\includegraphics[width=0.99\textwidth]{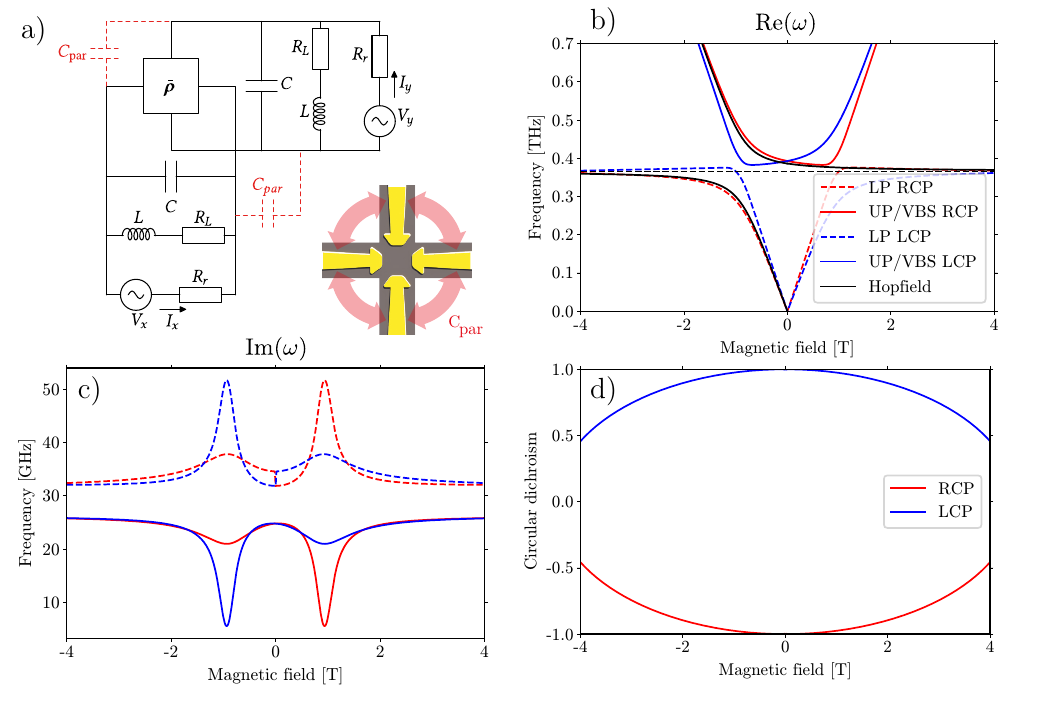}
	\caption{\label{fig:figparcap1} (a) Circuit model with the parasitic capacitance between the antennas. (b) Real and (c) imaginary parts of the dispersion obtained by solving Eq.~\ref{eq:diagcircuit}. The real part exhibits a smaller splitting of the counter-rotating mode. (d) Magnetic field-dependent circular dichroism of the eigenvectors. }
\end{figure*}

Close to the gap there is a significant parasitic capacitance between the orthogonal antennas, the effect of this can be considered by looking at the system in the diagonal basis (see Eq.~\ref{eq:diagcircuit}) obtained by rotating the coordinate system by \SI{45}{\degree}.
\begin{widetext}
	\begin{equation}\label{eq:diagcircuit}
		\begin{pmatrix}I_x - I_y \\ I_x + I_y\end{pmatrix} = \begin{pmatrix}\sigma_{xx} + \frac{1}{Z_C} + \frac{1}{Z_L} & -\sigma_{xy} + \frac{1}{Z_{C_{par}}} \\ \sigma_{xy} + \frac{1}{Z_{C_{par}}} & \sigma_{xx}+\frac{1}{Z_C} + \frac{1}{Z_L}\end{pmatrix}\begin{pmatrix}V_x - V_y \\ V_x + V_y\end{pmatrix}
	\end{equation}
	\begin{equation}
		\Longleftrightarrow \begin{pmatrix}I_x \\ I_y\end{pmatrix} = \begin{pmatrix}\sigma_{xx} + \frac{1}{Z_C} + \frac{1}{Z_L} - \frac{1}{Z_{C_{par}}}& -\sigma_{xy} \\ \sigma_{xy} & \sigma_{xx} + \frac{1}{Z_C} + \frac{1}{Z_L} + \frac{1}{Z_{C_{par}}}\end{pmatrix}\begin{pmatrix}V_x \\ V_y\end{pmatrix}
	\end{equation}
\end{widetext}
This represents two diagonal modes with the same inductance and capacitance as the original circuit, but coupled by the parasitic capacitance. They have a nominal resonance frequency of $\omega_\mathrm{cav} = \sqrt{\frac{1}{L(C-C_{par})}-\frac{R^2}{4L}}$. By diagonalizing we obtain modes satisfying
\begin{equation}
	\lambda_{\pm} = \sigma_{xx} - \sigma_{xy}e^{\pm i\phi} + \frac{1}{Z} - \frac{1}{Z_{C_{par}}} = 0
\end{equation}
with eigenvectors
\begin{equation}
	\vec{e}_{\pm} = \frac{1}{\sqrt{2}}\begin{pmatrix}1 \\ e^{\pm i \phi}\end{pmatrix},\quad \cos(\phi) = \frac{1}{Z_{C_{par}}\sigma_{xy}}
\end{equation}

In order to have well-defined modes we need $\Im(\cos(\phi)) = 0$ and this gives a condition for which the phase $\phi$ is independent of $\omega$
\begin{equation}
	\cos(\phi) = -\frac{2C_{par}(1+\tau^2\omega_c^2)}{\sigma_{DC}\tau^2\omega_c}
\end{equation}
Consequently each mode possesses a total of four different branches with a magnetic field-dependent phase $\phi$. It is evident from this formula that in an identical geometry ($C_{par}$ fixed), a sample with multiple quantum wells can be expected to exhibit less effects of the parasitic interaction since while $\tau$ might be smaller than for a structure with only a single quantum well, the increased electron density will more than compensate for this. 

\subsection{Model parameters}

The inductance of the antenna was calculated using a Drude model for the conductivity of gold $\sigma(\omega) = \frac{\sigma_{DC}}{1+i\omega\tau}$ with $\sigma_{DC} = \SI{5.9e28}{\siemens\per\metre}$ and $\tau=\SI{120}{\femto\second}$. Since the model contains a voltage source and not an electric field the effective inductance is \SI{33}{\pico\henry}. This corresponds to the inductance of a single nano-antenna times its length divided by two. The factor of two comes from each of the $\lambda/4$ nano-antennas in each direction being in parallel and not in series. The capacitance of an antenna array with a gap of \SI{2.5}{\micro\metre} was estimated to be \SI{7}{\nano\farad} and with a parasitic capacitance of approx. \SI{1.4}{\nano\farad}, which corresponds to $C_{par} \approx 0.2C$. It was obtained using the RLC solver of CST Microwave Studio. The radiation resistance was chosen to be $R_r = \SI{73}{\ohm}$, the input impedance of a centre-fed half-wave dipole antenna. 

The 2DEG for a single quantum well heterostructure had a nominal electron density of $n_{2D} = \SI{4.5e11}{cm^{-2}}$ and the Drude lifetime was assumed to be $\tau=\SI{15}{\pico\second}$. For the multiple quantum well structures a similar electron density was assumed for each quantum well and a dephasing time of $\tau=\SI{7.5}{\pico\second}$ was used, this is smaller than the transport Drude lifetime but consistent with the lifetime measured optically\cite{cyclotrondecaytime}. 

\begin{figure*}
	\includegraphics[width=0.99\textwidth]{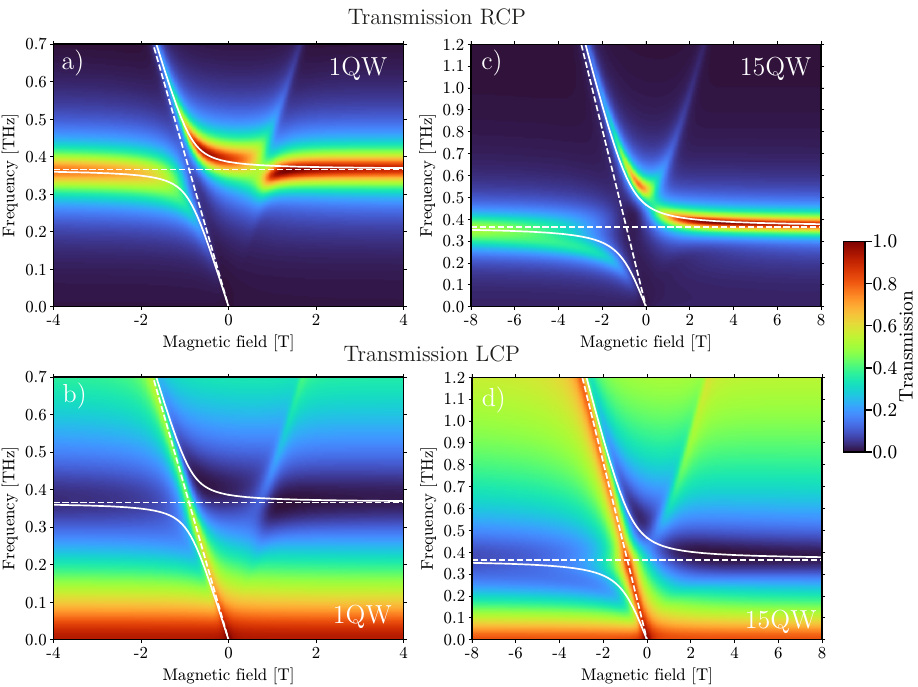}
	\caption{\label{fig:figparcapcolorplots} Circuit model with a parasitic capacitance of $C_{\mathrm{par}} = 0.2C$ for incoming RCP: (a) Transmission of RCP and (b) LCP for a single quantum well structure. (c) Transmission of RCP and (d) LCP for a 15 quantum well structure. }
\end{figure*}

In Fig.~\ref{fig:figparcap1} (b) the real part of the dispersion of a single quantum well structure is shown for the above parameters. There is an additional smaller splitting of the branch corresponding to the counter-rotating mode. While the splitting is small it `looks' larger in transmission because of the large increase in the imaginary part as seen in Fig.~\ref{fig:figparcap1} (c). The magnetic field-dependent circular dichroism is shown in Fig.~\ref{fig:figparcap1} (d). At higher magnetic fields the chirality is reduced, reducing the frequency difference between the VBS and LP modes. When the CD reaches zero these two modes merge. 

\section{Finite-element modelling}\label{finiteelement}

\begin{figure*}
	\includegraphics[width=1.0\textwidth]{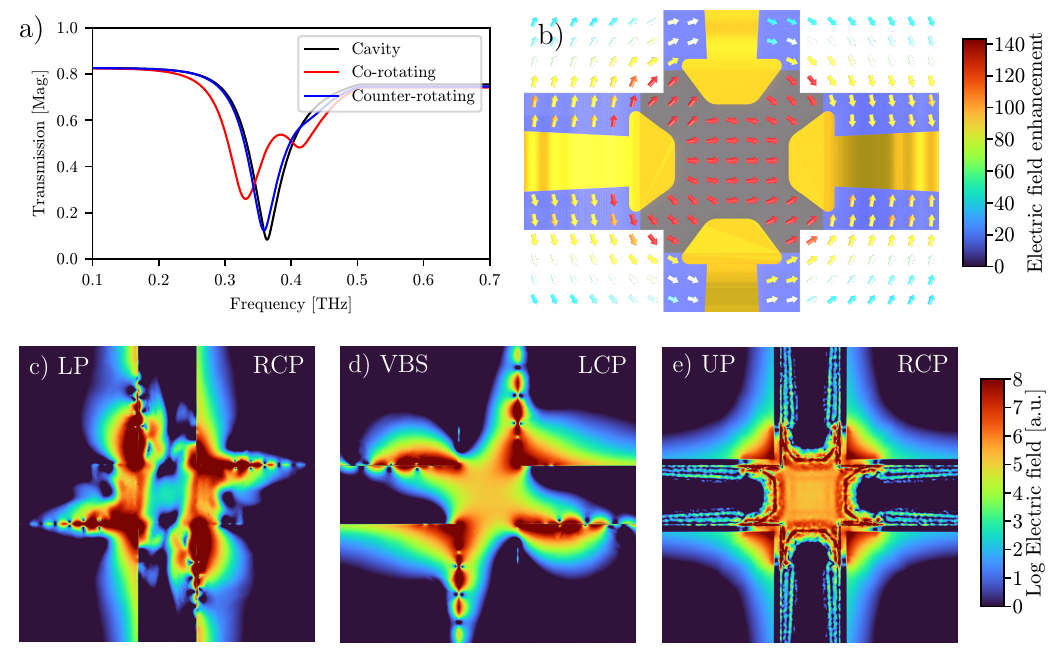}
	\caption{\label{fig:figfem}(a) Simulated transmission as function of frequency for $\omega_{c} = \omega_\mathrm{cav}$. (b) Electric field enhancement extracted from simulation for the empty cavity (without gyrotropic material). For $\omega_{c} = \omega_\mathrm{cav}$: (c) Electric field of the LP mode obtained from a simulation where the incoming light is coupling actively to the electron cyclotron motion, (d) electric field of the VBS mode where the incoming light is counter-rotating to the electron cyclotron motion and (e) the electric field of the UP mode, again with incoming light co-rotating with the electron cyclotron motion. }
\end{figure*}

The finite-element (FEM) simulations were done using the frequency domain solver of CST Microwave Studio with unit cell periodic boundary conditions along the $x$ and $y$ directions and with open boundaries along $z$. The 2DEG was simulated with a \SI{0.1}{\micro\metre} thick gyrotropic material. 

In Fig.~\ref{fig:figfem} (a) the simulated transmission can be found in the anti-crossing region where the cyclotron and cavity frequencies coincide. Fig.~\ref{fig:figfem} (b) shows the electric field enhancement, extracted by dividing the electric field distribution by the magnitude of the incoming electric field. The electric field enhancement consequently reaches approx.~$140$ inside of the gap. Fig.~\ref{fig:figfem} (c), (d) and (e) show the electric field distribution inside of the cap layer between the 2DEG and the antenna/\ce{SiO2} , showing the presence of excited propagating surface magneto-plasmon modes in the case of the UP mode and edge magneto-plasmon modes in the case of the LP and VBS modes. This can explain why mainly the dispersion of the LP and VBS modes is affected by surface magneto-plasmons, whereas for the UP mode it is the transmission.

\section{Heterostructures and fabrication}\label{appendix:fab}

\begin{figure*}
	\includegraphics[width=1.0\textwidth]{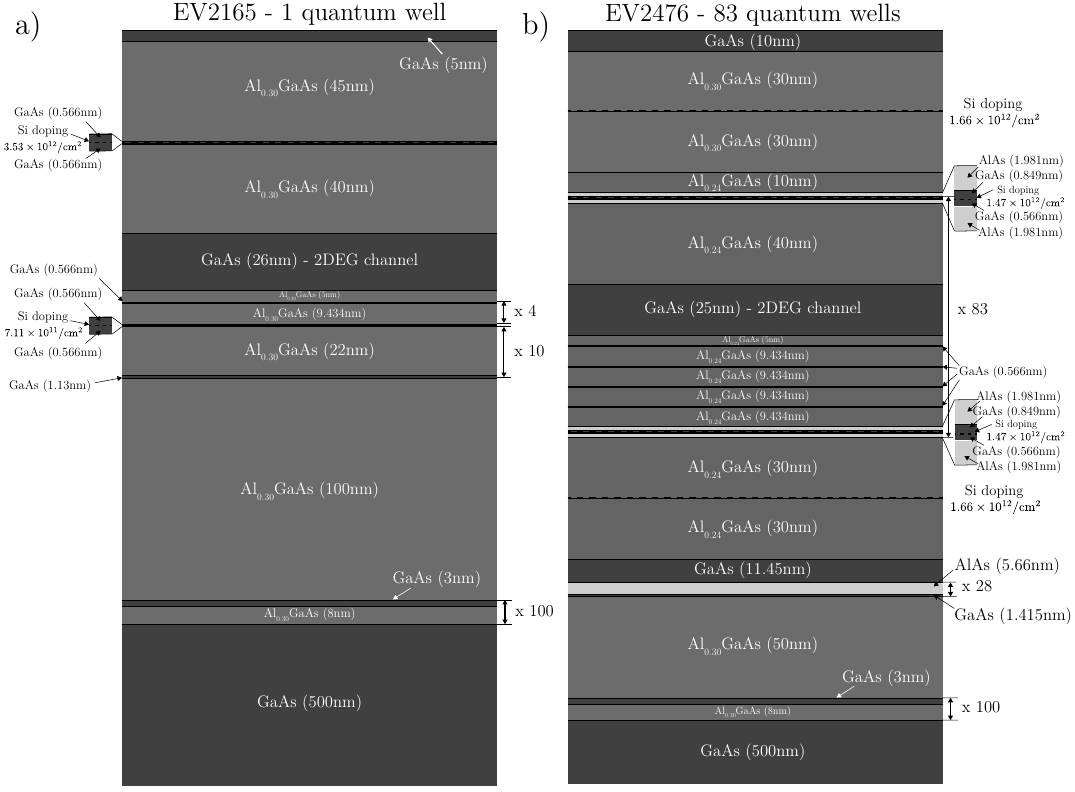}
	\caption{\label{fig:heterostructure}Schematics of the heterostructures used for the fabrication of (a) single quantum well samples and (b) multi-quantum well samples. }
\end{figure*}

The samples were fabricated using GaAs/AlGaAs heterostructures grown by MBE. A number of different AlGaAs/GaAs heterostructures were used for the fabrication. For the single quantum well samples a double-side doped heterostructure referred to as EV2165 was used with a nominal electron density of $\SI{4.5e11}{cm^{-2}}$, for the multiple quantum well samples a heterostructure with 83 QWs were used as a starting point with a similar electron density in each quantum well, these were then etched to the desired number of quantum wells by means of a slow sulphuric etch. 

In a first step four patches consisting of Ti/Au \SI{7}{\nano\metre}/\SI{200}{\nano\metre} were deposited surrounding each cross using e-beam. All photolithography was performed using direct laser writing. 

Afterwards a \SI{1}{\micro\metre} \ce{SiO2} layer was deposited using PECVD, the region of the patches was then etched using RIE and then a short etch using buffered HF was done in order to smoothen the sidewalls surrounding the patch in order to be able to deposit the antennas connecting the patches and to create a non-periodic modulation of the \ce{SiO2}  layer. Then Ti/Au \SI{10}{\nano\metre}/\SI{250}{\nano\metre} antennas were deposited using electron-beam deposition. After the antennas had been deposited a second layer of \ce{SiO2} approximately \SI{1.5}{\micro\metre} was deposited using PECVD to be used as a hard mask for etching away the 2DEG outside of the antennas and gap. 

Finally a deep etch of $\sim\SI{10}{\micro\metre}$ was done using ICP as a final step to reduce the amount of cross-talk between the orthogonal sets of antennas outside of the 2DEG. An approximately \SI{20}{\nano\metre} thin \ce{Al2O3} layer was deposited to passivate the side-walls and prevent oxidation of the quantum wells. 

\section{THz time-domain magneto-spectroscopy measurement with linear THz polarization}\label{appendix:tdslinpol}

\begin{figure*}
	\includegraphics[width=1.0\textwidth]{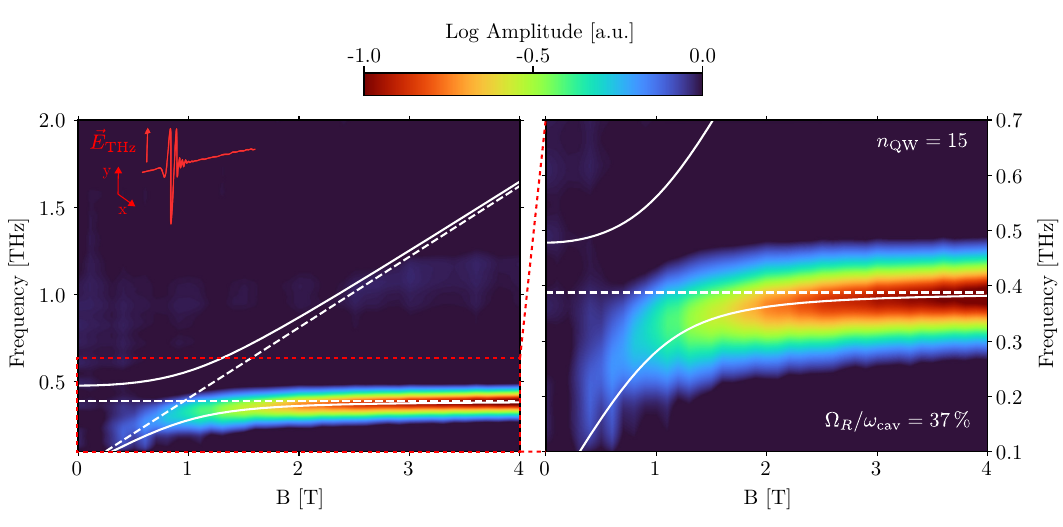}
	\caption{\label{fig:tdslinpol15QW}Terahertz time-domain spectroscopy measurement of the sample with 15 quantum wells with vertical THz polarization together with a fit using a single-mode Hopfield model\cite{hagenmuller2010ultrastrong}.}
\end{figure*}

The transmission measurement of linearly polarized THz radiation shown in Fig.~\ref{fig:tdslinpol15QW} was performed without the two terahertz retarders and without any additional optical elements between the photoconductive switch used to generate THz pulse and the ZnTe crystal employed for electro-optic detection.

%\bibliography{../references}

\end{document}